\documentclass[12pt]{article}

\usepackage{epsfig}

\usepackage{scicite}

\usepackage{times}

\newcommand\apss{{\it Astrophys. Space Sci.}}
\newcommand\ApJ{{\it Astroph. J.}}
\newcommand\apj{{\it Astroph. J.}}
\newcommand\aj{{\it Astron. J.}}

\newcommand\mnras{{\it Mon. Not. Royal Astron. Soc.}}

\newcommand\AsA{{\it Astron. Astroph.}}
\newcommand\etal{{\it et al.\ }}
\newcommand{\dm}{\Delta m_{15}}
\normalsize

\newenvironment{sciabstract}{%
\begin{quote} \bf}
{\end{quote}}

\newcounter{lastnote}
\newenvironment{scilastnote}{%
\setcounter{lastnote}{\value{enumiv}}%
\addtocounter{lastnote}{+1}%
\begin{list}%
{\arabic{lastnote}.}
{\setlength{\leftmargin}{.22in}}
{\setlength{\labelsep}{.5em}}}
{\end{list}}

\title{
Spectropolarimetric diagnostics of thermonuclear supernova
explosions} 

\author
{Lifan Wang$^{1,2}$ \footnote{To Whom correspondence should be addressed. E-mail: wang@physics.tamu.edu}, Dietrich Baade,$^{3}$ Ferdinando Patat$^{3}$\\
\\
\normalsize{$^{1}$Physics Department, Texas A\&M University, College Station, 
Texas, 77843-4242}\\
\normalsize{$^2$Lawrence Berkeley National Laboratory, 1 Cyclotron Rd, Berkeley, CA 94710}\\
\normalsize{$^{3}$European Southern Observatory,Karl-Schwarzschild-Strasse 2,D-85748 Garching, Germany}
}

\date{}

\begin{document} 


\baselineskip24pt


\maketitle


\begin{sciabstract}
Even at extragalactic distances, the shape of supernova 
ejecta can be effectively diagnosed by 
spectropolarimetry. 
We present here results for 17 Type Ia supernovae that 
allow a statistical study of the correlation among
the geometric structures and other observable parameters
of Type Ia supernovae.
These observations suggest that their ejecta typically 
consist of a smooth, central iron rich core and an outer layer with 
chemical asymmetries. 
The degree of this peripheral asphericity 
is correlated with the light-curve decline rate of Type Ia supernovae.
These observations lend strong support to delayed-detonation models
of Type Ia supernovae.
\end{sciabstract}



Different supernova explosion mechanism may lead to differently structured ejecta.
Type Ia supernovae (SNe) have been used as a premier tool for 
precision cosmology. 
They occur when a carbon/oxygen white 
dwarf reaches the Chandrasekhar stability limit, probably 
due to mass accretion in a binary system, and is disrupted 
in a thermonuclear explosion\cite{Whelan:1973}. It is, 
however, a matter of decade-long debate how the 
explosive nuclear burning is triggered and how it propagates 
through the progenitor star
\cite{Arnett:1969,Nomoto:1976,Khokhlov:1991,Reineke:2002,Plewa:2004}. 
Successful models generally start with a phase of sub-sonic 
nuclear burning, or deflagration, but theorists disagree on 
whether the burning front becomes supersonic following the 
earlier phase of deflagration. An explosion that does turn 
into supersonic burning is called delayed-detonation\cite{Khokhlov:1991}. 
The resulting chemical structures are dramatically different for
deflagration\cite{Reineke:2002} and delayed-detonation
models\cite{Gamezo:2005}. For a pure deflagration model, chemical clumps
are expected to be
present at all velocity layers that burning has reached\cite{Reineke:2002}. For 
delayed-detonation, the detonation front propagates through 
the ashes left behind by deflagration and burns 
partially burned or unburned elements further into heavier 
elements, and erases the chemically clumpy structures generated 
by deflagration. 

The polarized emission from a supernova is caused by electron
scattering in its ejecta. It is sensitive to the
geometric structure of the ejecta\cite{Hoeflich:91}. 
Electron
scattering in an asymmetric ejecta would produce non-zero degrees of polarization\cite{Hoeflich:91}.
In continuum light, normal 
SNe~Ia are only polarized up to about 0.3\%, but 
polarization as high as 2\% is found across some 
spectral 
lines\cite{Wang:1996,Wang:1997,Howell:01,Wang:03a,Wang:2004,Leonard:2005}.  
The polarization 
decreases with time
 and
vanishes around 2 weeks past optical maximum. The low level of 
continuum polarization implies that the
SN~Ia photospheres are, in general, approximately spherical. 
The decrease of
polarization at later times suggests that the outmost zones of the 
ejecta are more aspherical than the inner zones.
Similarly, the large polarization across certain spectral lines 
implies that the layers above the photosphere are highly 
aspheric and most likely 
chemically clumpy\cite{Wang:03a,Wang:2004,Kasen:03,Leonard:2005}.

In this study, we report polarimetry of 17 SNe~Ia. These are all the SNe Ia
for which we have pre-maximum polarimetry. The
observations (Table 1) were collected with the 2.1-m Otto Struve Telescope of
the McDonald Observatory of the University of Texas and with one of
the 8.2-m unit telescopes of Very Large Telescope (VLT) of the 
European Southern Observatory (ESO). The typical spectral resolution of
these observations is 1.2 nm and the wavelength range is typically from 400nm to 850nm. 
The total exposure time is  
longer than 4 hours for the data taken with the 2.1-m telescope, and is 
around 1-2 hours for the ESO-VLT. 
Only SNe observed before optical maximum are included. 
We restrict ourselves to 
the characteristic Si\,II 635.5\,nm line. More information about data acquisition and reduction is given in the
Supporting Online Material.

The observed 
Stokes parameters can be projected onto the so-called principle and 
secondary axes, which are
defined\cite{Wang:01,Wang:03a} by a principle-component analysis of
the data points on the Q-U diagram such that the spectral variation of the
polarization is maximal along the principle axis. The secondary axis
is orthogonal to the principal axis.
As an example, in the spectra and principle components of 
the polarization of SN~2002bo at different epochs (Fig.~1),
the polarized
spectral features at 470.0\,nm, and at 600.0\,nm 
decreased significantly by day +14. 
The polarization data for SN~1996X, SN~1999by, SN~2001el, 
and SN~2004dt have been discussed in previous 
studies\cite{Wang:1997,Howell:01,Wang:03a,Kasen:03,Wang:2004,Leonard:2005} 
revealing a considerable level of individuality.

\begin{figure}[thb]
\psfig{figure=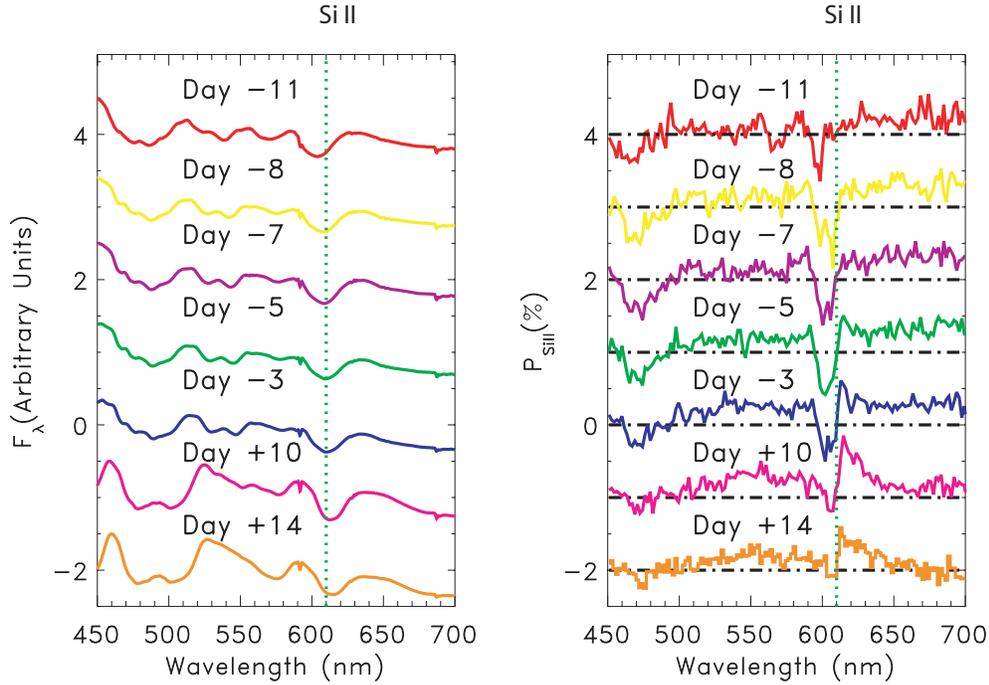,width=13cm,clip=}
\caption{Spectroscopy (left) and spectropolarimetry (right) of SN~2002bo. 
The units of the fluxes on the left panel are arbitrary,
from top to bottom, the zero point of each spectrum is offset from zero by
4, 3, 2, 1, 0, -1, and -2 for clarity. On the right hand side panel, from 
top to bottom, the degree of polarization of each observation is offset by 
4\%, 3\%, 2\%, 1\%, 0\%, -1\%, and -2\%. In both panels, the vertical 
dotted line mark the absorption features of the Si\,II lines.}
\end{figure}

As a luminosity indicator we use the decline in $B$ 
magnitude within 15 days from maximum ($\dm$). This 
quantity is found to be well correlated with the intrinsic luminosity of
SNe~Ia\cite{Phillips:1993}: intrinsically dimmer SNe~Ia
usually show a faster decline, i.e.\ larger $\dm$.   
For those SNe with no published light curves, the 
luminosity indicator was derived from 
cross correlations of the flux spectra with a library of 
well-observed SNe~Ia, and the $\dm$ values of the closest 
spectral matches were adopted. The best matches to SN~2002fk, 
SN~2003W, SN~2004ef, SN~2005de, and SN~2005df are
SN~1996X, SN~2002bo, 
SN~2001el, SN~2001el, and SN~1994D, 
respectively. The results are cross-checked with the ratio
of the Si\,II absorption lines at 615 and 580 nm \cite{Nugent:1995},
and the results are generally in good agreement.
For the SNe with no published light curves, 
0.05 mag is quadratically added to the errors
of $\dm$ deduced from the light curve of the best 
matching spectral template. All $\dm$\ values  
derived from published light curves are based on our light-curve 
fitting procedure\cite{Wang:2006}.  

The degree of polarization is known to evolve with time. 
To compare the data at the same epoch, the time dependence of the
polarization of all the Branch normal SNe is
fitted by a second order polynomial (shown in Figure~S1) 
and the polarization data are then corrected to 5 days before peak $B$ magnitude. 
Excluding peculiar events
SN~2001V, SN~2004dt, and SN~1999by, (see Supporting Online Material), 
the correlation is fitted by a linear relation $P_{SiII}\ = \
0.48(03)+1.33(15)(\Delta m_{15}-1.1)$.  
The Pearson correlation 
coefficient is 0.87 which suggests a strong correlation. 
The $\chi^2$ of the linear fit is 16 with 13 degrees of freedom.

\begin{figure}
\psfig{figure=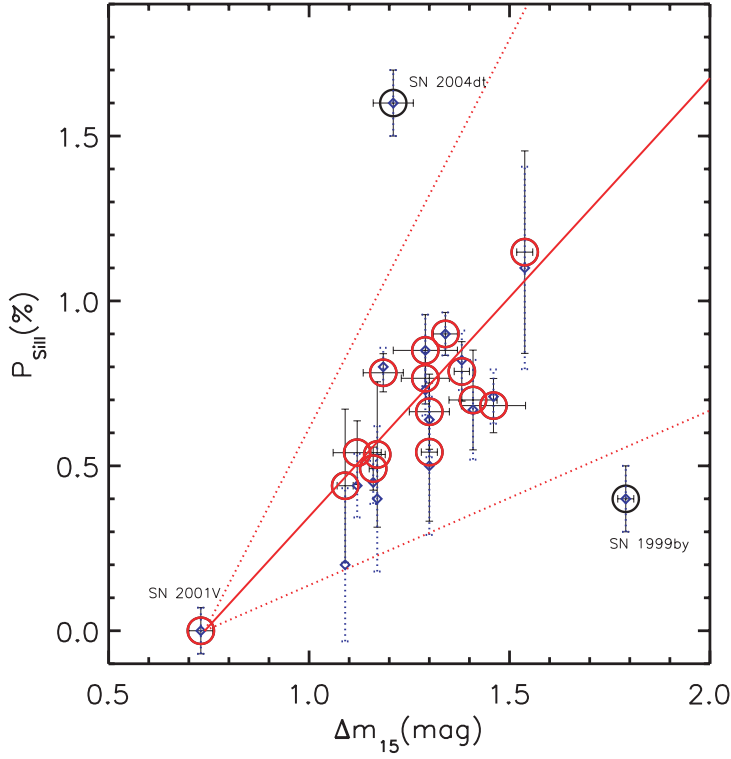,width=10cm,height=10cm,clip=}
{\caption{The correlation between the degree of polarization
across the Si\,II 635.5 nm line and the light curve decline rate
$\dm$\ for a sample of 17 Type-Ia SNe.
The blue diamonds are the measured values as given in 
Table~1. The open circles show the data corrected to day -5, by subtracting
$0.041(t+5)+0.013(t+5)^2$ from the observed degree of 
polarization, where $t$ is the day after optical maximum.  This correction formula was derived from a weighted 
quadratic fit to the time dependence of polarization (shown in the insert). 
The linear fit represented by the straight
line includes only spectroscopically normal SNe shown as red open circles. The blue
open circle shows the highly polarized event SN~2004dt, and the sub-luminous SN~1999by. The dotted-lines
illustrate the 1-$\sigma$ level of the intrinsic polarization distribution
 around the most-likely
value for the Monte-Carlo simulation described in the
text. 
}}
\end{figure}

As the continuum polarization is generally low ($\sim 0.2$) for SN~Ia, the ejecta is likely to be
chemically clumpy to explain the observed high polarization across spectral lines\cite{Wang:2004}.
If this is the case, projection effects must play a role in diluting 
this correlation (Fig.~2). This might be the case 
for the significant deviation of SN~2004dt from the regression 
line (Fig.~2). When viewed in certain directions, the
degree of polarization can be particularly large or small.
A tight correlation could only be expected for the mean 
degree of polarization averaged 
over several SNe and at the same photometric phase. 
To quantify this effect, we have constructed a toy-model which assumed
that the disk of the supernova photosphere is polarized at 
11.73\%\cite{Chandra:1960} at the limb, and decreases 
quadratically with distance from the limb to the center
to zero polarization (see also the Supporting Online Material). We performed a Monte-Carlo 
simulation which assumes $N$ opaque clumps, each covering an area  $S$ along the
line of sight to the photosphere. The depth of the
P-Cygni absorption feature 
(defined as the ratio of flux at the minimum of the absorption feature
to that of the continuum) is then $1-NS/\pi R^2$, where $R$ is the radius of
the photosphere. For lines about 50\% deep 
($NS/\pi R^2\ =\ 0.5$), the probability distribution of the
polarization is found to peak at
around 0.5\% for $N\ = \ 20$. The $1-\sigma$ width of this distribution 
calculated assuming $NS/\pi R^2\ =\ 0.5$ (Fig.~2) envelopes most of the 
observed data points. In reality, there is a wide range of  Si II line strengths, accordingly the numbers and sizes  of lumps  may not be the same for all the SNe. The observed polarization-$\Delta m_{15}$ correlation appears to be tighter than given by the Monte-Carlo model. This is perhaps an indication of a non-negligible amount of large scale asymmetry of the SN ejecta, especially for those SNe with high polarization and $\Delta m_{15}$.  Such large scale asymmetries do not generate noticeable amounts of polarization in the continuum, which is formed deeper inside, and can be due to large plumes located above the SN photosphere. It may also be generated from the interaction of the ejecta with the circumstellar material such as an accretion disk before the explosion of the white dwarf progenitor.
Alternatively, the observed tight correlation may also be due to a global aspherical explosion. In this case, a tight correlation implies more asymmetric explosion generates intrinsically dimmer SNe.
However, we stress that the asymmetries we observed here are confined to the high velocity regions and do not affect the geometric shape of SN photosphere around optical maximum. Any large scale asymmetry is therefore confined only to the outmost layers. 


As the light curve of SN~Ia is 
powered by the radioactive decay of $^{56}$Ni, we infer 
(Fig.~2) a possible anti-correlation between 
the amount of $^{56}$Ni synthesized in SNe~Ia and the
asphericity of the silicon rich layer. 

Our discovery puts strong constraints on any successful 
models of SNe~Ia. 
At around optical maximum, the photosphere is typically located at
velocities around 12,000 km/sec as measured from P-cygni line profiles, 
which according to hydrodynamic
calculations\cite{Gamezo:2005} of delayed-detonation, is 
close to the velocity zones dominated by iron group elements.
The absence of significant
polarization at this velocity is evidence in support of
delayed-detonation.

Details of delayed-detonation models affect the
brightness of SNe~Ia and their geometric structure.  
Larger departures from sphericity imply
less of the central region is scoured of irregularities in the 
composition left by pure deflagration models, and thus
less material burned to
thermonuclear equilibrium and hence dimmer SNe, 
in accordance with the statistical trend revealed by our
studies. 


Finally, we make some remarks on using SNe~Ia as standard candles. Asymmetry
introduces intrinsic magnitude and color dispersions.  
Intrinsic color dispersion may be particularly important as it makes it
difficult to perform precise extinction corrections. The stochastic nature of
the origin of the asymmetry suggest that the  color corrections can only 
be performed in a statistical sense. It
is perhaps difficult to
find pairs of SNe~Ia with identical light-curve and spectroscopic 
properties. 

In summary, the application of spectropolarimetric observing
techniques to SNe~Ia permits the geometric structures of SNe to be
probed even though they are at distances that cannot be spatially resolved.
The explosion of SNe~Ia is intrinsically a 3-D
phenomenon, and a phase of delayed detonation is necessary to
account for the observed geometric and chemical differentiation.

\bibliography{scibib}

\bibliographystyle{Science}

{}


\begin{scilastnote}
\item {We are grateful to the European Southern Observatory for the generous
allocation of observing time. We especially thank the staff of the
Paranal Observatory for their competent and never-tiring support of
this project in service mode.  The research of LW is supported in part
by the Director, Office of Science, Office of High Energy and Nuclear
Physics, of the U.S. Department of Energy under Contract
No. DE-AC03-76SF000098.  We are grateful to discussions with J.\ Craig
Wheeler.  The SN polarimetry project, on which this study is
based, has greatly benefited from contributions by him and Peter
H\"oflich. We are also grateful to the anonymous referees that helped to 
improve this paper.
This work is based in part on observations obtained at the
European Southern Observatory, Chile (Program IDs:  64.H-0617(B),  66.D-0328(A),  
67.D-0517(A),  68.D-0571(A),  69.D-0438(A), 70.D-0111(A),  71.D-0141(A),  
073.D-0771(A), and 075.D-0628(A))}
\end{scilastnote}


\clearpage

\begin{table}
\caption{The identification, observing date, phase (in days relative to 
optical maximum light), telescope, intrinsic degree of polarization
across the Si II 635.5 nm line, $\Delta\ m_{15}$, and references for $\dm$ values}
\begin{tabular}{l|lll|lll}
\hline
SN & Date of Obs & Phase & Telescope & P$_{SiII}$ & $\Delta m_{15}$ & References\\
\hline
1996X & 1996/04/14  & -4.2 & McD 2.1 m  & 0.50(20)   & 1.30(02)   & \cite{Riess:99,Wang:2006}\\
1997bp & 1997/04/07 &-5.0 & McD 2.7 m  & 0.90(10) & 1.29(08)   & \cite{Jha:2006,Wang:2006}\\
1997bq & 1997/04/25 &-3.0 & McD 2.1 m  & 0.40(20)   & 1.17(02) & \cite{Jha:2006,Wang:2006}\\
1997br & 1997/04/20 &-2.0 & McD 2.1 m  & 0.20(20) & 1.09(02)   & \cite{Li:99,Wang:2006}\\
1999by & 1999/05/09 &-2.5 & McD 2.1 m  & 0.40(10)   & 1.79(01) & \cite{Garnavich:99}\\
2001V & 2001/02/25 &-7.3  & VLT        &  0.00(07)& 0.73(03)   & \cite{Vinko:03,Wang:2006}\\
2001el & 2001/09/26& -4.2 & VLT &  0.45(02)  & 1.16(01)  & \cite{KRISCIUNAS:03,Wang:2006}\\
2002bo & 2002/03/18 &-5.0 & VLT &  0.90(05)   & 1.34(03)   & \cite{Benetti:04,Wang:2006} \\
2002el & 2002/08/14 &-6.4 & VLT &  0.72(09)   & 1.38(05)   & \cite{Wang:2006} \\
2002fk & 2002/10/06 &-5.5 & VLT &  0.67(10)   & 1.19(05)   & see text \\
2003W & 2003/01/31 &-4.5  & VLT &  0.64(10)   & 1.30(05)   & see text \\
2004dt & 2004/08/13 &-7.3 & VLT &  1.60(10)   & 1.21(05)   & \cite{Pastorello:2006} \\
2004ef & 2004/09/11 & -4.1 & VLT & 1.10(30)   & 1.54(07) & see text\\
2004eo & 2004/09/20 & -5.9 & VLT & 0.71(08)   & 1.46(08) & \cite{Pastorello:2006}\\
2005cf & 2005/06/01 & -9.9 & VLT & 0.44(05)  & 1.12(06) & \cite{Pastorello:2006}\\
2005de & 2005/08/06 & -4.4 & VLT & 0.67(14) & 1.41(06) & see text \\
2005df & 2005/08/08 & -4.3 & VLT & 0.73(05) & 1.29(09) & see text\\
\hline
\end{tabular}
\end{table}
\end{document}